\documentclass[10pt,preprint,a4paper]{aastex}

\usepackage{graphics,epsf}
\usepackage{amsmath}                
\usepackage{amsfonts}               
\usepackage{amssymb}                
\usepackage{epsfig}                 

\def \cm{~\rm{cm}}
\def \s{~\rm{s}}
\def \km{~\rm{km}}

\def \g{~\rm{g}}

\def \AU{~\rm{AU}}
\def \erg{~\rm{erg}}

\def \yr{~\rm{yr}}

\title{FORMATION OF BIPOLAR PLANETARY NEBULAE BY INTERMEDIATE-LUMINOSITY OPTICAL TRANSIENTS}

\author{Noam Soker\altaffilmark{1} and Amit Kashi\altaffilmark{1}}

\altaffiltext{1}{Department of Physics, Technion -- Israel Institute of
Technology, Haifa 32000 Israel; kashia@physics.technion.ac.il; soker@physics.technion.ac.il.}

\begin{document}

\begin{abstract}
We present surprising similarities between some bipolar planetary nebulae (PNe) and eruptive
objects with peak luminosity between novae and supernovae.
The later group is termed ILOT for intermediate luminosity optical transients (other terms are
intermediate luminosity red transients and red novae).
In particular we compare the PN NGC~6302 and the pre-PNe OH231.8+4.2, M1-92 and IRAS~22036+5306
with the ILOT NGC~300~OT2008-1.
These similarities lead us to propose that the lobes of some (but not all) PNe and pre-PNe
were formed in an ILOT event (or several close sub-events).
We suggest that in both types of objects the several months long outbursts are powered by mass
accretion onto a main-sequence companion from an AGB (or extreme-AGB) star.
Jets launched by an accretion disk around the main-sequence companion shape the bipolar lobes.
Some of the predictions that result from our comparison is that the ejecta of some ILOTs will
have morphologies similar to those of bipolar PNe, and that the central stars of the PNe that
were shaped by ILOTs should have a main-sequence binary companion with an eccentric orbit of
several years long period.
\end{abstract}

\section{INTRODUCTION}
\label{sec:intro}

Peak supernovae (SNe) luminosities are about four orders of magnitude above those of novae.
This gap is slowly filled with observations of eruptive events
(e.g.,
\citealt{Barbary2009}; \citealt{Berger2009},\citeyear{Berger2011}; \citealt{Bond2009}; \citealt{Kulkarni2007}; \citealt{KulkarniKasliwal2009};
\citealt{Kasliwal2011}; \citealt{Ofek2008}; \citealt{Rau2007}; \citealt{Mould1990}; \citealt{Mason2010}; \citealt{Pastorello2010};
\citealt{Prieto2008}, \citeyear{Prieto2009}; \citealt{Botticella2009}; \citealt{Smith2009}).
We term these outbursts Intermediate-Luminosity Optical Transients (ILOTs; other terms in use are
Intermediate-Luminosity Red Transients and Red Novae).

The pre-outburst objects of some of the ILOTs were identified to be asymptotic giant branch (AGB) or extreme-AGB stars,
with NGC~300 OT2008-1 (\citealt{Bond2009}; hereafter NGC~300OT) being the prototype.
{{{  If the AGB star looses substantial amount of mass in such an event (but not necessarily its entire envelope),
and the event takes place not too long before the AGB star turns to a
planetary nebula (PN), then the descendent PN is expected to have the following characteristic properties: }}}
\begin{enumerate}
\item A linear velocity-distance relation.
The ILOTs lasts for a time period $\Delta t_I$ of weeks to several years.
When the ejected mass is observed at a time $t_{\rm PN} \gg \Delta t_I$
(hundreds to tens thousands of years) later, each mass element is at a distance of its velocity times $t_{\rm PN}$.
Therefore, the PN component that was ejected during the ILOT is expected to posses a linear relation between
velocity and distance.
Slowing down elements will have a velocity lower than the fastest parts of the nebula having a similar distance from the center.
{{{  Other PN elements that were lost before or after the event will not share this velocity to distance linear relation.
It might be hard to tell whether slower elements come from an earlier mass loss episode or were part of the ILOT event but
have been slown down. }}}
\item Bipolar structure.
As the ILOT is expected to result from a binary interaction (\citealt{KashiSoker2010b}; \citealt{SokerKashi2011}),
the PN component ejected during the ILOT is expected to have a bipolar structure.
For example, two lobes, or point-symmetrical structure if the jets that are lunched during the ILOT precess.
{{{  In that respect, most PNs that have been formed by an ILOT event, and hence are bipolar,
are expected to host a binary system with an orbital separation of $\sim 1 \AU$.
Another possibility is that the ILOT event took place just as the companion entered the common envelope. }}}
\item Expansion velocity of few$\times 100 \km \s^{-1}$.
As we think that most ILOTs are powered by accretion onto a main-sequence (MS) star
(\citealt{KashiSoker2010b}; \citealt{SokerKashi2011}) that blows jets, the {{{  maximum }}} outflow velocity is
similar to that of the escape velocity from MS stars.
{{{  The fastest moving elements will be dense parcels of gas that were only slightly slowed-down
by the interaction with the AGB wind.
The average velocity of the ejecta will be several times lower than the escape velocity from the companion
because of the interaction with the slower AGB wind, but still much faster than the AGB wind velocity. }}}
Therefore, the faster parts of the PN component that was ejected by an ILOT are expected to
move at velocities of $\sim 100$ -- $1000 \km \s^{-1}$.
\item Total kinetic energy of $\sim 10^{46}$ -- $10^{49} \erg$.
As typical ILOT energy is in that range, we expect the kinetic energy of the ejected component to be in that range.
\end{enumerate}

In this paper we argue that some bipolar PNe that have components showing these four properties
might have been formed in an ILOT event.
When such a PN is observed, we have no information on the exact duration of the mass ejection event.
If the event was too long, then even if the total energy is as expected, the luminosity might have been
too low (below typical novae luminosity) and the event was not an ILOT.
Bearing this in mind, we nonetheless go ahead and compare, in section \ref{sec:6302}, the
PN NGC~6302 with the ILOT NGC~300OT.
We note that \citet{Prieto2009} already made a connection between the ILOT NGC~300OT and pre-PNe.
Based on that they raised the possibility that the progenitor of NGC~300OT was of mass $< 8 ~\rm{M_\odot}$.
In section \ref{sec:others} we discuss three pre-PNe that we suggest formed by ILOTs.
Motivated by the similarities between the ILOT and these PNe, in section \ref{sec:scenarios} we
discuss plausible scenarios for the formation of a PN in an ILOT.
Our discussion and summary are in section \ref{sec:summary}.

\section{COMPARING THE ILOT NGC~300OT WITH THE PN NGC~6302}
\label{sec:6302}

The basic characteristics of the PN NGC~6302 are summarized most recently by \citet{Szyszka2011}.
The relevant properties are summarized in Table \ref{Table:PN}.
The fundamental property that motivated us to make the comparison with an ILOT is the velocity-distance
linear relation, $v_r \propto r$, that points out to a short lobes-ejection event (\citealt{Meaburn2008}; \citealt{Szyszka2011}).
It is important to emphasize that some components of the nebula do not obey this relation,
implying they were ejected over a relatively long time before or after the lobes-ejection event.
Such is the dense massive torus of mass $\sim 2~\rm{M_\odot}$ (\citealt{Matsuura2005}; \citealt{Peretto2007}; \citealt{Wright2011};
We note that \citealt{Dinh-V-Trung2008} obtained that the mass of the torus is only $0.087~\rm{M_\odot}$).
The torus was ejected over $\sim 5000 \yr$ prior to the lobes-ejection event \citep{Peretto2007}.
\begin{table}
\caption{Comparing PN and ILOT.}

\bigskip

\small
\begin{tabular}{|l|c|c|c|}
 \hline
                                         & NGC~6302$^1$            & NGC~300OT$^2$                         & OH231.8+4.2$^4$                    \\
 \hline
Type of object                           & PN                        & ILOT                                & pre-PN                             \\
 \hline
Mass source                              & AGB star                  & Extreme-AGB star                    & Mira star (AGB)                    \\
 \hline
Early mass loss$^3$:                     &  Equatorial torus         &                                     &                                    \\
 Mass loss rate ($\rm{M_\odot} \yr^{-1}$)& $\sim 5 \times 10^{-4}$   & $\sim 6 \times 10^{-4}$             &                                    \\
 Ejection Velocity ($\km \s^{-1}$)       & $\sim 10$                 & $\sim 12$                           &                                    \\
 Duration (yr)                           & $\sim 5000$               & $<10^4$                             &                                    \\
 Total mass ($\rm{M_\odot}$)             & $\sim 2$                  & $\sim 5$                            &                                    \\
Dust                                     & $\sim 0.03~\rm{M_\odot}$  & Like in pre-PNe$^7$                 &                                    \\
\hline
Eruption:                                &                           &                                     &                                    \\
 Duration (yr)                           & $\lesssim 100$            & $\sim0.22$                          & $< 125$                            \\
 Ejected mass ($\rm{M_\odot}$)           & $0.1$ -- $1$              & $\sim 0.5$                          & $\sim 0.3$                         \\
 Velocity range ($\km \s^{-1}$)          & $0$ -- $500$              & $75$ -- $1000$                      & $0$ -- $\ga 400$                   \\
 \hline
Total energy$^5$ (erg)                   & $0.4$ -- $2\times 10^{47}$ & $\sim 2$ -- $10 \times 10^{47}$    & $\sim 3 \times 10^{46}$            \\
\hline
Stellar Mass ($\rm{M_\odot}$)            & $M_1 \simeq 6 $            & $M_1 \sim 6$ -- $15$               & $M_1 \simeq 3.5$                   \\
on the ZAMS$^6$                          &                            &                                    & $M_2 \simeq 2$                     \\
\hline
Prediction                               & A solar-like               & 1) Bipolar ejecta                  &  1) The companion                  \\
($P$ orbital period;                     & MS companion               & from the ILOT.                     &  orbit has $e \ga 0.3$             \\
$e$ eccentricity)                        & at $\sim 2$ -- $5 \AU$,    & 2) A $\sim 3$ -- $10~\rm{M_\odot}$ &  and $P \sim 3$ -- $10 \yr $       \\
                                         & possibly in an             & MS companion.                      &  2) The system might go            \\
                                         & eccentric orbit:           & Orbit: $e \ga 0.3$;                &  through an ILOT event             \\
                                         & $e \ga 0.3$;               & $P \sim 5$ -- $50 \yr$             &  in the near future.               \\
                                         &  $P \sim 3$ -- $10 \yr $   &                                    &                                    \\
\hline
\end{tabular}
\footnotesize

\bigskip
(1) Data for NGC~6302 are from \cite{Meaburn2008}, \cite{Szyszka2011}, \cite{Matsuura2005}, \cite{Wright2011}.
\newline
(2) Data for NGC~300OT from \cite{Kochanek2011}, \cite{Prieto2009}, \cite{Bond2009}, \cite{Kashi2010}.
\newline
(3) The mass loss episode prior to the eruption.
\newline
(4) Parameters for OH231.8+4.2 are from \citeauthor{Kastner1992} (\citeyear{Kastner1992},\citeyear{Kastner1998}), \citeauthor{Contreras2002} (\citeyear{Contreras2002},\citeyear{Contreras2004}), \cite{Alcolea2001}, \cite{Bujarrabal2002}.
\newline
(5) Large fraction of the mass in lobes of PNe moves at low velocity.
\newline
(6) Zero age main-sequence.
\newline
(7)  The optical similarity to pre-PNs is discussed in \citet{Prieto2009}.
\label{Table:PN}
\end{table}

 \citet{Prieto2009} make the connection between the ILOTs NGC~300OT and SN~2008S to pre-PNe, and suggest similar
progenitors. \citet{Prieto2009} based their conclusions on the similarities of the mid-IR spectrum, optical spectra,
kinematics, and dusty circumstellar medium. By kinematics they refer to the expansion velocity and bipolar morphology.
We here add the similar properties of total energy and short ejection event, and discuss the
formation of the bipolar structure in PNe and pre-PNe by ILOT events.

\begin{figure*}
\resizebox{1.0\textwidth}{!}{\includegraphics{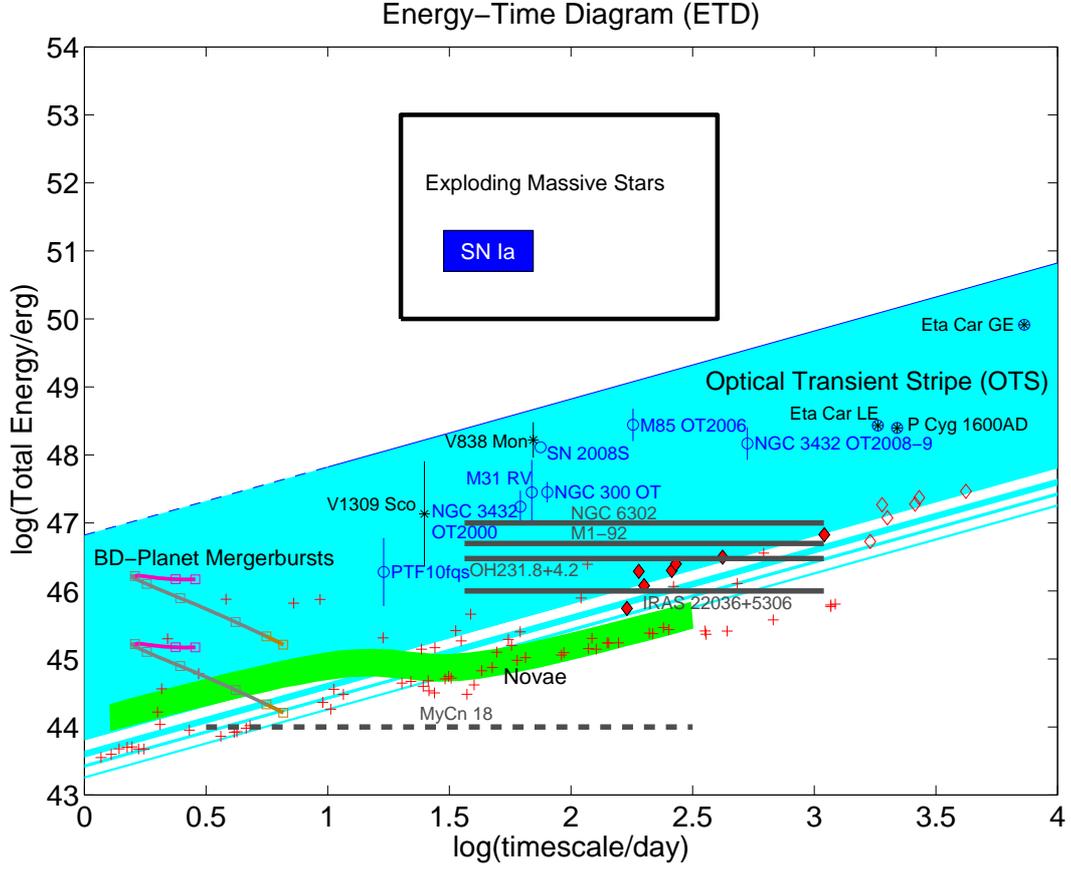}}
\caption{\footnotesize{Observed transient events on the Energy-Time Diagram (ETD).
Blue empty circles represent the total (radiated plus kinetic) energy
of the observed transients as a function of the duration $t$ of their eruptions.
The Optical Transient Stripe (OTS), is a more or less constant luminosity region in the ETD.
It is populated by accretion powered events such as ILOTs (including mergerbursts),
major LBV eruptions, and predicted BD-planets mergerbursts \citep{Bear2011}.
The green line represents nova models computed using luminosity and duration from \cite{dellaValleLivio1995}.
Nova models from \cite{Yaron2005} are marked with red crosses, and models from Shara et al. (2011) are represented with diamonds.
The total energy does not include the energy which is deposited in lifting the envelope that does not escape from the star.
Where a model exists to calculate the gravitational energy released by the accreted mass (the available energy),
it is marked by a black asterisk.
The kinetic energies of the components that expand with a linear velocity-distance relation in each of the
five bipolar planetary nebulae and pre-PNe discussed in this paper, are marked by horizontal lines.
The energy of each object is derived from observations (for uncertainties see Table \ref{Table:PN}),
while the time scale is an estimate of our proposed model.
MyCn~18 is an exception, as it was formed by a nova eruption rather than by an ILOT.}}
\label{fig:totEvst}
\end{figure*}

The kinetic energy of the gas in the lobes suggests to us that the lobes-ejection event was of the
same magnitude as ILOT events.
To demonstrate this we draw the kinetic energy of the lobes on the
Energy-Time diagram (ETD) that is used to characterized ILOTs (\citealt{Kashi2010}; \citealt{KashiSoker2010b}).
The ETD (Figure \ref{fig:totEvst}) presents the \emph{total energy} of the transients, radiated plus kinetic,
as a function of the duration of their eruption, defined as a drop of $3$ magnitudes in the V-band.
When there is more information on a transient from observations or modelling, we present in the ETD the
\emph{available energy}, i.e., total gravitational energy available for the event.
Namely, the gravitational energy that could have been released if all the mass is accreted by the accreting star.
However, for most ILOTs the observations and models are not yet detailed enough to perform this estimate,
and we can only present the estimated radiated plus kinetic energy.

The upper-right region of the Optical Transient Stripe (OTS) is occupied by observed major luminous blue variable (LBV) eruptions.
For major LBV eruptions the available energy is equal to the radiated plus kinetic energy, as there is no
inflated envelope.
The observed lower-left region is occupied by ILOTs.
There are no observed objects yet in further lower-left of the OTS, but this region is predicted to be
occupied by brown dwarf (BD)-planet mergerbursts \citep{Bear2011}.
In this process the planet is shredded into a disk, and the accretion lead to an outburst.
The destruction of a component in a binary system and transforming it to an accretion disk is an extreme
case of mass transfer processes in binary systems.

The ILOT NGC~300OT was discovered by \cite{Monard2008},
with a bolometric luminosity of $L_{\rm bol} = 1.6 \times 10^{40} \erg \s^{-1}$ at discovery \citep{Bond2009}.
The pre-outburst progenitor was later reported by \cite{Prieto2008}.
 Spectra taken by \emph{SPITZER} revealed that most of the energy was emitted in the IR (\citealt{Prieto2009}).
It was enshrouded by dust (\citealt{Bond2009}; \citealt{Berger2009}),
and had a luminosity of about $6 \times 10^4 ~\rm{L_\odot}$, corresponding to a
$M = 10$ -- $15~\rm{M_\odot}$ extreme-AGB star (\citealt{Thompson2009}; \citealt{Botticella2009}).
A more massive red supergiant of mass $M = 12 - 25~\rm{M_\odot}$, as suggested by
\cite{Gogarten2009} based on stellar evolution considerations, may also be consistent with the data.
 \citet{Prieto2009} noticed the similarity of NGC~300OT to pre-PNe and put the lower mass range on $\sim 6 ~\rm{M_\odot}$.
They also raised the possibility that the progenitor can be a C-rich AGB star.
\cite{Kochanek2011}, on the other hand, favors a $9~\rm{M_\odot}$ progenitor.

\cite{Berger2009} attributed the $\sim 10^3 \km \s^{-1}$ red wing of the
Ca II H\&K absorption lines either to an infalling gas from a previous
eruption or to a wind of a companion star.
In either case the star accreting this matter is likely to be a MS star.
Together with the evidence for the supergiant nature of the progenitor, this
implies that there are two different stars in the system.
\cite{Bond2009} interpreted the Hydrogen Balmer lines and the Ca~II IR triplet's double features
as indicating the presence of a bipolar outflow expanding at a velocity of $\sim 75 \km \s^{-1}$.
\cite{Bond2009} suggested that NGC~300OT originated from an evolved massive star
on a blue loop to warmer temperatures, and was subjected to an increased instability due to prior mass loss.

\cite{Patat2010} observed an asymmetric dusty environment extending a few-thousand$\AU$
surrounding NGC~300OT that hints to a previous possible eruption.
This asymmetry may further hint to the presence of a companion star,
although we note that up to date there has been no definite observation proving a companion existence.
\cite{Thompson2009} suggested that ILOTs occur due to single
star processes, e.g., electron-capture SN, an explosive birth of a massive WD,
or an enormous outburst of a massive star.
In their model for NGC~300OT like events,
the progenitors are luminous ($\sim 4$ -- $6 \times 10^4~\rm{L_\odot}$) dust-enshrouded stars,
at the end of their AGB stage.

In \cite{Kashi2010} we suggested that NGC~300OT was powered by a mass transfer event of
few$\times 0.1~\rm{M_\odot}$ from an extreme-AGB star to a $3-10~\rm{M_\odot}$ MS companion.
One of the arguments for the presence of a companion is that a faster observed outflow of up to $\sim 600 \km \s^{-1}$
(\citealt{Berger2009}) fits better the escape velocity from MS stars.
{{{ The total energy of the eruption (radiated and kinetic), $E_{\rm{tot}} \simeq 2$ -- $4 \times 10^{47} \erg$, was therefore explained as having a gravitational origin. }}}

There is no problem in our lack of information about the amount of energy that was radiated during
the lobes-ejection event of NGC~6302.
The reason is that in ILOTs the kinetic energy is typically much larger than the radiated energy.
For NGC~300OT, for example, estimates of the kinetic energy are in the order of
few$\times 10^{47} \erg$ (\citealt{KashiSoker2010b}; \citealt{Kochanek2011}), much higher than the radiated energy.

We also note that our claim for an ILOT event does not imply that there was only one event.
It is possible, as in the case of the LBV in NGC~3432 (also referred to as SN~2000ch), that several
short events occurred one after the other.
This LBV underwent a major eruption in 2000 which lasted 62 days \citep{Wagner2004}, followed by
a series of three eruptions in 2008-2009, lasting for a total of $\sim 531$ days \citep{Pastorello2010}.

Just as a plausible example for the PN NGC~6302, there could have been, say, 10 short sub-events, each with an
energy of few$\times 10^{46} \erg$, lasting for $\sim 30-100~$day, and repeating
periodically, due to an eccentric orbit every $\sim 3$ -- $10 \yr$.
Each such sub-event falls well within the OTS.
Therefore, in Figure \ref{fig:totEvst} we mark the time scale for the events as we estimate based on our ILOT model,
$\sim 0.1$ -- $3 \yr$.
This estimate well satisfies the part of the orbit where the companion was close to the primary,
and could have triggered the events.

{{{  We note that the formation of a bipolar nebula via mass loss from an evolved star in a close binary system
does not necessarily imply a decrease in eccentricity.
This is seen observationally for $\eta$ Car and the Red Rectangle,
a bipolar nebula around a post-AGB star in a binary system with an orbital period of 322~days and an eccentricity of $e=0.34$.
A possible explanation to the high eccentricity of the Red Rectangle was worked out in \cite{Soker2000a}.
There it was shown that enhanced mass loss rate during periastron passages can overcome tidal effects, and the
eccentricity might increase even.
In \cite{KashiSoker2010a} we found that during the Great Eruption of $\eta$ Car the eccentricity almost
did not change, despite large amounts of mass loss and mass transfer.
The eccentricity of the $\eta$ Car binary system was very high before the Great Eruption.
Namely, despite the close approach at periastron, the eccentricity of the system was very large, $e \sim 0.9$,
before the Great Eruption, and stayed large after the eruption.
Over all, both observations and theory support the notion that the binary systems discussed in this study
can maintain high eccentricities. }}}

\section{OTHER PLANETARY NEBULAE AND PROTO-PNe}
\label{sec:others}

There are other bipolar PNe and pre-PNe that might have been formed by ILOTs (in one event or several sub-events).
We consider here three pre-PN as further examples.
 More pre-PNe were compared to NGC~300OT by \citet{Prieto2009}.
We end this section by presenting one bipolar PN that was not formed by an ILOT despite having a linear
velocity-distance relation.

\subsection{The bipolar pre-PN OH231.8+4.2}
\label{subsec:OH231}

OH231.8+4.2 (The Calabash Nebula, also known as ``The Rotten Egg Nebula'')
is a bipolar pre-PN (e.g., \citealt{Bujarrabal2002}) that we suggest was formed by an ILOT.
The central system of the pre-PN comprises a bright Mira variable star with an estimated zero age MS (ZAMS) mass of
$\sim 3.5~\rm{M_\odot}$ (QX Pup; \citealt{Kastner1992},\citeyear{Kastner1998}),
and an A star companion \citep{Contreras2004}.
According to \cite{Alcolea2001} and \cite{Bujarrabal2002} the lobes reach a velocity of $\ga 400 \km s^{-1}$,
have a momentum of $\sim 3 \times 10^{39} \g \cm \s^{-1}$, a total kinetic energy of $\sim 3 \times 10^{46} \erg$,
and were formed over a time scale of $< 125 \yr$.

\citet{Contreras2004} suggest that the lobes were inflated by jets in the velocity range $v_j = 500-1000 \km \s^{-1}$.
This gives a total jets' mass of $M_{j} \simeq 0.01$ -- $0.02~\rm{M_\odot}$, and total kinetic energy of
$E_j \simeq 0.5$ -- $1 \times 10^{47} \erg$.
These jets interacted with the circumbinary gas and inflated the lobes that now contain $\sim 0.3~\rm{M_\odot}$
(both lobes contain similar masses in spite of their different extents).
This interaction dissipated part of the kinetic energy of the jets.

As with NGC~6302, there are components that were not ejected or were not accelerated during the lobes-ejection event.
Most of the nebular mass, $\sim 0.64~\rm{M_\odot}$, does not reside in the lobes, but rather resides near the center
and expands at low velocities $< 40 \km \s^{-1}$ \citep{Alcolea2001}.
Some mass is in a torus of radius $\sim 6 \AU$ very close to the center \citep{Contreras2002}.

What is most interesting about OH231.8+4.2 is that the primary star is still an AGB star, and
that it has a MS companion of spectral type A0 and a mass of $\sim 2~\rm{M_\odot}$ \citep{Contreras2004}.
This is the type of a companion that is required in the binary ILOT model that we have developed over the
last several years \citep{KashiSoker2010b}.
It is very likely (although not necessary) that the companion has an eccentric orbit,
and with an orbital period of $\sim 3$ -- $10 \yr$.

\citet{Alcolea2001} and \citet{Contreras2004} considered the active jets phase to have last $\sim 100 \yr$.
\citet{Contreras2004} proposed that the eruption during these hundred years was like that of FU~Ori type outbursts of young stars.
We instead suggest that this phase, the lobes-ejection event, was much shorter. It was composed from one ILOT event
or several close ILOT sub-events, each lasting for $\la 100~$days.
In Figure \ref{fig:totEvst} we mark the time scale as we estimate based on our ILOT model,
$\sim 0.1$ -- $3 \yr \ll 100 \yr$.
As the primary of OH231.8+4.2 is still an AGB star, such an event might take place again.
Namely, according to our model it is quite possible that OH231.8+4.2 will experience an ILOT event in the near future.

\subsection{The bipolar pre-PN M1-92}
\label{subsec:PN192}
Another example is the dusty \citep{Ueta2007} bipolar (e.g., \citealt{Trammell1996}) pre-PN M1-92 (IRAS~19343+2926)
that has a general linear velocity-distance relation \citep{Bujarrabal1998b}, and an extra kinetic energy of
$\sim 7 \times 10^{45} \erg$ \citep{Bujarrabal1998a}.
By extra kinetic energy \cite{Bujarrabal1998a} refer to the energy of the lobes above that of the regular AGB wind.
This extra energy, they suggest, comes from the post-AGB jets \citep{Bujarrabal1998a}.
The extra energy is calculated from the velocity component along the symmetry axis of the lobes, that reaches a maximum
velocity of $\sim 70 \km \s^{-1}$.

The kinetic energy in the event could have been higher, but it was dissipated when the ejected jets where interacting with
the more massive circumbinary gas.
The extra-momentum along the axis is $\sim 3 \times 10^{39} \g \cm \s^{-1}$ \citep{Bujarrabal1998a}.
If the ejected speed of the jets was $v_j > 70 \km \s^{-1}$, i.e., more than
the maximum observed present speed in the lobes, then the ejected mass was $<0.2~\rm{M_\odot} (v_j/70 \km \s^{-1})^{-1}$.
The corresponding kinetic energy is $E_j \simeq 10^{46} (v_j/70 \km \s^{-1}) \erg$.
As a more typical value we take the ejected velocity to be that of an escape speed from a low mass MS star,
$\sim 300$ -- $400 \km \s^{-1}$, for which we get $E_j \simeq 5 \times 10^{46} \erg$, and $0.04~\rm{M_\odot}$
for the ejected mass in the jets.

The general linear velocity-distance relation, the maximum present lobes' velocity of $\sim 70 \km \s^{-1}$, and the total energy
suggests to us that the lobes were shaped by jets \citep{Bujarrabal1998a}, that were ejected in an
ILOT event (or several close sub-events).
As we did for the previous objects, we mark on Figure \ref{fig:totEvst} the timescale as we estimate it
from our ILOT model, $\sim 0.1$ -- $3 \yr$.

\subsection{The bipolar pre-PN IRAS~22036+5306}
\label{subsec:I22036}
The pre-PN IRAS~22036+5306 (hereafter I22036) shows that the ejection of a bipolar component can last
for $\la 1 \yr$.
\cite{Sahai2003} presented \emph{HST} observations of this pre-PN, revealing an extended knotty bipolar shape.
The structure of I22036 is rather complicated and has many sub features \citep{Sahai2003}.
The total mass of the pre-PN is $\sim 0.065~\rm{M_\odot}$ and the velocities of the various components range between
$0$ -- $450 \km \s^{-1}$ \citep{Sahai2006}.

\cite{Sahai2006} presented spectra of the object, revealing an interesting fast ($v \la 220 \km \s^{-1}$) bipolar molecular outflow,
which erupted in $\sim 1981$.
The mass in this component of I22036 was estimated to be $\sim 0.03~\rm{M_\odot}$, giving a kinetic energy of $\sim 10^{46} \erg$.
The fast molecular outflow component obeys a linear velocity law up to an outer bow-shock region, implying that the ejection event was
much shorter than the age at observation that was $\sim 25 \yr$.
In Figure \ref{fig:totEvst} we mark the time scale as we estimate based on our ILOT model, $0.1$ -- $3 \yr \ll 25 \yr$.
Another common property of this and NGC~300OT and the two previous pre-PNe is that the AGB progenitor was massive, $M_{\rm ZAMS} \ga 5 ~\rm{M_\odot}$.
As we discuss in section \ref{sec:scenarios}, such massive AGB stars are likely to suffer instabilities that can cause them to lose
a large portion of their envelope in a very short time.

The short ejection time, the linear velocity-distance relation, the bipolar structure, the massive progenitor, and the outflow velocity range
raise the possibility that the ejection event was an ILOT.
This leads us to predict that the central star of I22036 has a MS companion of mass $\sim 1 ~\rm{M_\odot}$ with an eccentric orbit
with a period of few years.

How come an ILOT in the year 1981 was not observed? We note that the central star of I22036 is heavily obscured \citep{Sahai2003}.
We therefore suggest that the ILOT heated the circumbinary dust, and most radiation came out in the IR before IRAS was launched, hence
avoided detection as an outburst.

It is interesting to mention another indirect evidence for a companion \citep{Sahai2011b}.
The very broad H$\alpha$ wings observed in I22036 \citep{Sahai2006}, can be interpreted as
Raman scattering of Ly$\beta$ arising in the ionized gas region observed by EVLA \citep{Sahai2011a}, which is surrounded by a neutral region.
The required width of Ly$\beta$ ($\sim 400 \km \s^{-1}$) might be generated in an accretion disk around a companion \citep{Sahai2011b}.

\subsection{A counter example}
\label{subsec:counter}
Not all bipolar PNe and pre-PNe were formed/shaped in an ILOT event.
A linear velocity-distance relation implies short-period formation event, but not necessarily an ILOT,
as is the case with the Hourglass Nebula (MyCn~18).
The knots along the polar directions, that are at larger distances than the hourglass structure,
have a linear velocity-distance relation. The maximum velocity is $\sim 500 \km \s^{-1}$ \citep{O'Connor2000}.
Although these properties are as in the other systems discussed here, the total mass of the knots is much
smaller, and amounts to $\la 10^{-4}~\rm{M_\odot}$ \citep{Sahai1999}.
\citet{O'Connor2000} estimate the total mass in the knots to be $\sim 10^{-5}~\rm{M_\odot}$ and their velocity
in the range of several hundreds to over $600 \km \s^{-1}$.
Over all, the kinetic energy of the knots is $\la 10^{44} \erg$, fitting nova outbursts.
Indeed, \citet{O'Connor2000} suggest that the knots were formed from a nova outburst.
We agree with this assessment, and hence don't consider the knots in the
hourglass nebula to have been formed in an ILOT event,

\section{A PLAUSIBLE SCENARIO}
\label{sec:scenarios}

We propose that the lobes of NGC~6302, OH231.8+4.2, M1-92, I22036, and similar objects with similar bipolar morphologies,
a similar lobes' kinetic energy range, and an expansion velocity that shows a linear velocity-distance relation, are formed in an ILOT event.
Our model for ILOTs (\citealt{SokerKashi2011}; \citealt{KashiSoker2010b}) is a mass transfer onto a MS (or slightly evolved) star.
The mass transfer can be in one of two basic processes.
In the first process a merger process occurs. A low mass star is destructed on the MS star, as is the mergerburst model for
V838~Mon (\citealt{TylendaSoker2006}; \citealt{SokerTylenda2006}). This is not the case considered here.

In the second process an evolved giant star (LBV, AGB, extreme-AGB) enters an unstable phase.
The interaction with a companion causes the star to lose a huge amount of mass in a very short time.
Part of this mass is accreted by the MS companion via an accretion disk.
The accretion disk launches two jets that form the lobes.
The process is accompanied by high luminosity that makes the event an ILOT.
An extreme example of such a process is the 20 years long Great Eruption of $\eta$ Car where the lobes are thought to have been shaped
by jets launched by the mass-accreting companion (\citealt{Soker2001}; \citealt{KashiSoker2010a}).
In \cite{Kashi2010} we have already made the connection between the ILOT NGC~300OT and the Great Eruption of $\eta$ Car.
A connection between $\eta$ Car and a nebula around a post-AGB star -- The Red Rectangle -- was conducted in \citet{Soker2007}.
Here we make a direct connection between ILOTs and the formation process of the lobes in some
(but not all) bipolar PNe and pre-PNe.
 An earlier comparison of NGC~300OT progenitor to pre-PNs was made by \citet{Prieto2009} based
on optical and kinematical properties.

{{{  There is a question whether the very high mass accretion rate envisioned in our model can lead to the formation of jets.
An encouragement for a positive answer are FU Orionis (FU Ori) outbursts.
These are Sun-like protostars (YSO) that undergo a rapid accretion episode.
The typical mass accretion rate is $\sim 10^{-4} ~\rm{M_\odot} \yr^{-1}$ and mass outflow rate of $\sim 10\%$ the accretion rate
(e.g., \citealt{Reipurth2002}).
\cite{Hartmann2011} report on a YSO of $0.3~\rm{M_\odot}$ accreting an a rate of $\sim 2 \times 10^{-4} M_\odot \yr^{-1}$.
In their theoretical study \cite{Baraffe2010} take protostars of $\sim 0.1~\rm{M_\odot}$ to
accrete at a rate of $5 \times 10^{-4} ~\rm{M_\odot} \yr^{-1}$.
Therefore, it is quite possible that main sequence stars of $\sim 1$ -- $5~\rm{M_\odot}$ can accrete mass at a very high rate as required in our
proposed scenario.
The physics of jets launching in FU Ori outbursts might be the same as in YSO objects with much lower accretion rates \citep{Konigl2011}.
Magnetic fields that are required in launching jets can be amplified by a dynamo operating in the accretion disk.
In any case, the accreted mass must get rid of most of its angular momentum, and close to the stellar surface jets can efficiently
carry the extra angular momentum. }}}

The 20 years long eruption of $\eta$ Car had 4 spikes in its light curve.
It is also possible that in the systems studied here the interaction occurs over several orbital periods,
with mass accretion and jet launching occurring only at periastron passages.
In $\eta$ Car the kinetic energy is $\sim 1000$ larger than in the present systems,
and the companion has a mass of $\sim 30$ -- $80~\rm{M_\odot}$
(e.g., \citealt{Mehner2010}; \citealt{KashiSoker2010a} and references therein).
The mass accreted onto the companion during the great eruption is several$\times \rm{M_\odot}$
(\citealt{Soker2001}, \citeyear{Soker2007}; \citealt{KashiSoker2010a}).
MS stars of mass $\sim 0.3$ -- $3 ~\rm{M_\odot}$ have a gravitational potential well similar to that of the Sun.
To explain a jets' power of $E_j$ the accreted mass onto a solar like MS star should be
\begin{equation}
M_{\rm acc} \simeq \frac{2 E_j \rm{R_\odot}} {G \rm{M_\odot}} = 0.05 \left( \frac{E_j}{10^{47} \erg} \right) ~\rm{M_\odot} . 
\label{eq:macc}
\end{equation}

In the case of NGC~6302 the amount of mass in the lobes is estimated to be $0.1$ -- $2.5~\rm{M_\odot}$.
The upper limit is given in a recent analysis by \citet{Wright2011}.
However, we suspect that most of this mass is closer to the center and moves at a low velocity.
The ejected mass in the lobes that appear in Table \ref{Table:PN} is the higher velocity gas that has the linear velocity-distance relation
\citep{Szyszka2011}, and for that we take an average velocity of $\sim 200 \km \s^{-1}$ \citep{Szyszka2011}.
Over all, the kinetic energy in the lobes is highly uncertain, and we take it to be $0.4$ -- $2 \times 10^{47} \erg$.
Allowing for $\sim 50$ per cent efficiency of the process where the jets accelerated and inflated the lobes, requires the jets' energy to be
$1$ -- $5 \times 10^{47} \erg$.
The companion had to accreted $0.05$ -- $ 0.25~\rm{M_\odot}$.
If the jets were launched at $\sim 700 \km \s^{-1}$, their mass amounts to $0.02$ -- $0.1~\rm{M_\odot}$.
Namely, the mass lost in the jets is $\sim 20$ -- $40$ per cent of the mass transferred to the companion.
This is similar to the fraction in the model for the Great Eruption of $\eta$ Car \citep{KashiSoker2010a}.

Based on this discussion, we consider the following scenario.
First we note that the massive ($\sim 2.2~\rm{M_\odot}$; \citealt{Wright2011}) equatorial disk of NGC~6302 was formed over a time period of
$\sim 5000 \yr$, that ceased $\sim 650 \yr$ before the lobes-ejection event \citep{Szyszka2011}.
We propose that during this time there was a strong tidal interaction that lead to the formation of the equatorial mass loss process.
The binary system was stable against the Darwin instability, no Roche lobe overflow took place, and the system avoided a common envelope phase.
The orbital separation was about several AU.
After losing $\sim 2.2~\rm{M_\odot}$ from its envelope, the AGB star entered a more stable phase.
For example, its radius decreases.

After another 650 years the AGB entered another unstable phase. For example, a shell helium flash caused its
envelope to substantially expand. Else, a strong magnetic activity, as was suggested for the unstable phase
of the primary of $\eta$ Car during the Great Eruption \citep{Harpaz2009}, took place.
After all, the AGB star has a strong convection, and its envelope is spun-up by the tidal interaction with the companion.
With a very strong convection, even a slow rotation can make the AGB star magnetically active,
and it might experience a magnetic activity variation, even cyclical (\citealt{Soker2000b}; \citealt{Garcia2001}).
As a result of the radius increase a very strong tidal interaction took place, and a RLOF occurred.
This process is more pronounced if the orbit is highly eccentric, and the process takes place when the
companion approaches periastron.
This is the case in $\eta$ Car.
During the RLOF an accretion disk was formed, and two jets were launched.
This leads us to predict that the central star of NGC~6302 has a MS companion of mass
${\rm few}~\rm{M_\odot}$ in an eccentric orbit and an orbital period of several years.


\section{DISCUSSION AND SUMMARY}
\label{sec:summary}

 We presented surprising similarities between two seemingly unrelated groups of objects:
planetary nebulae (PNe) and eruptive objects with peak luminosity between novae and supernovae.
The later group is termed ILOT for intermediate luminosity optical transients (also termed
intermediate luminosity red transients and red novae).
The similarities between the ILOT NGC~300OT and the PN NGC~6302 are discussed in section \ref{sec:6302},
and similarities with two pre-PNe are discussed in section \ref{sec:others}.
They are summarized in Table \ref{Table:PN}.
A connection between the ILOT NGC~300OT and pre-PNe was made by \citet{Prieto2009}.

Basically, the lobes of these PNe and prot-PNe have kinetic energy similar to that of some ILOTs
(Fig. \ref{fig:totEvst}).
They also have a linear velocity-distance relation that points to a short ejection event, and
a velocity range that suggests a mass ejection by a main-sequence (MS) companion to the AGB progenitor.
In a case of low ejected mass and energy, as in the PN MyCn~18 (section \ref{subsec:counter}), a nova
outburst rather than an ILOT event launched the jets.

We suggest that the lobes of these PNe and pre-PNe were formed in one event or
several sub-events that occurred at periastron passages of the binary system.
{{{  We emphasize that not all binary progenitors of PNs experience an ILOT.
An ILOT event requires that the AGB suffers a major instability.
This probably requires a massive AGB star.
Also, the orbit should be highly eccentric to prevent a continues high-mass loss rate.
These conditions require further study. }}}
In our model that is summarized in section \ref{sec:scenarios} each such event might last for several months.
The AGB enters an unstable phase and loses a large amount of mass. Part of this mass
is accreted by the companion, and an accretion disk is formed. The accretion disk launches two jets
that inflate the lobes.
Our model is compatible with the MS companion of the pre-PN OH231.8+4.2.
The primary star in OH231.8+4.2 is still an AGB star (Mira variable), and we predict that another ILOT event is
possible in this system.
This and some other predictions of our model are listed in Table \ref{Table:PN}.
In particular, we predict that the ejected mass of ILOTs will possess a bipolar structure.
 We note that a large fraction of the outburst radiation of NGC~300OT was in the IR bands (\citealt{Prieto2009}),
and predict that in many cases ILOTs will be observed from AGB stars with close MS companions.

\bigskip

We thank Adam Frankowski, Raghvendra Sahai and Albert Zijlstra for helpful comments,
{{{  and an anonymous referee whose comments helped in improving the manuscript. }}}
AK acknowledges a grant from the Irwin and Joan Jacobs fund at the Technion.
This research was supported by the Asher Fund for Space Research at the Technion, and the Israel Science Foundation.


\footnotesize

\begin{thebibliography}

\bibitem[Alcolea et al.(2001)]{Alcolea2001} Alcolea, J., Bujarrabal, V., S{\'a}nchez Contreras, C., Neri, R., \& Zweigle, J.\ 2001, \aap, 373, 932

\bibitem[Baraffe \& Chabrier(2010)]{Baraffe2010} {{{  Baraffe, I., \& Chabrier, G.\ 2010, \aap, 521, A44 }}}

\bibitem[Barbary et al.(2009)]{Barbary2009} Barbary, K., et al. 2009, \apj, 690, 1358

\bibitem[Bear et al.(2011)]{Bear2011} Bear, E., Kashi, A., \& Soker, N. 2011 , \mnras (arXiv:1104.4106)

\bibitem[Berger et al.(2009)]{Berger2009} Berger, E., et al. 2009, \apj, 699, 1850

\bibitem[Berger et al.(2011)]{Berger2011}Berger, E., Foley, R., \& Soderberg, A. 2011, The Astronomer's Telegram, 3467

\bibitem[Bond et al.(2009)]{Bond2009} Bond, H.~E., Bedin, L.~R., Bonanos, A.~Z., Humphreys, R.~M., Monard, L.~A.~G.~B., Prieto, J.~L., \& Walter, F.~M.\ 2009, \apjl, 695, L154

\bibitem[Botticella et al.(2009)]{Botticella2009} Botticella, M.~T., et al. 2009, \mnras, 398, 1041

\bibitem[Bujarrabal et al.(1998a)]{Bujarrabal1998a} Bujarrabal, V., Alcolea, J., \& Neri, R.\ 1998, \apj, 504, 915

\bibitem[Bujarrabal et al.(1998b)]{Bujarrabal1998b} Bujarrabal, V., Alcolea, J., Sahai, R., Zamorano, J., \& Zijlstra, A.~A.\ 1998, \aap, 331, 361

\bibitem[Bujarrabal et al.(2002)]{Bujarrabal2002} Bujarrabal, V., Alcolea, J., S{\'a}nchez Contreras, C., \& Sahai, R.\ 2002, \aap, 389, 271

\bibitem[della Valle \& Livio(1995)]{dellaValleLivio1995} della Valle, M., \& Livio, M.\ 1995, \apj, 452, 704

\bibitem[Dinh-V-Trung et al.(2008)]{Dinh-V-Trung2008} Dinh-V-Trung, Bujarrabal, V., Castro-Carrizo, A., Lim, J., \& Kwok, S.\ 2008, \apj, 673, 934

\bibitem[Garc{\'{\i}}a-Segura et al.(2001)]{Garcia2001} Garc{\'{\i}}a-Segura, G., L{\'o}pez, J.~A., \& Franco, J.\ 2001, \apj, 560, 928

\bibitem[Gogarten et al.(2009)]{Gogarten2009} Gogarten, S.~M., Dalcanton, J.~J., Murphy, J.~W., Williams, B.~F., Gilbert, K., \& Dolphin, A.\ 2009, \apj, 703, 300

\bibitem[Harpaz \& Soker(2009)]{Harpaz2009} Harpaz, A., \& Soker, N.\ 2009, \na, 14, 539

\bibitem[Hartmann et al.(2011)]{Hartmann2011} {{{  Hartmann, L., Zhu, Z., \& Calvet, N.\ 2011, arXiv:1106.3343 }}}

\bibitem[Kastner et al.(1992)]{Kastner1992} Kastner, J.~H., Weintraub, D.~A., Zuckerman, B., Becklin, E.~E., McLean, I., \& Gatley, I.\ 1992, \apj, 398, 552

\bibitem[Kastner et al.(1998)]{Kastner1998} Kastner, J.~H., Weintraub, D.~A., Merrill, K.~M., \& Gatley, I.\ 1998, \aj, 116, 1412

\bibitem[Kashi et al.(2010)]{Kashi2010} Kashi, A., Frankowski, A., \& Soker, N.\ 2010, \apjl, 709, L11

\bibitem[Kashi \& Soker(2010a)]{KashiSoker2010a} Kashi, A., \& Soker, N.\ 2010a, \apj, 723, 602

\bibitem[Kashi \& Soker(2010b)]{KashiSoker2010b} Kashi, A., \& Soker, N.\ 2010b, (arXiv:1011.1222)

\bibitem[Kasliwal et al.(2011)]{Kasliwal2011} Kasliwal, M.~M., et al. 2011, \apj, 730, 134

\bibitem[Kochanek(2011)]{Kochanek2011} Kochanek, C.~S.\ 2011, ApJ, in press (arXiv:1106.4722)

\bibitem[K{\"o}nigl et al.(2011)]{Konigl2011} {{{  K{\"o}nigl, A., Romanova, M.~M., \& Lovelace, R.~V.~E.\ 2011, \mnras, 416, 757 }}}

\bibitem[Kulkarni et al.(2007)]{Kulkarni2007} Kulkarni, S.~R., et al. 2007, \nat, 447, 458

\bibitem[Kulkarni \& Kasliwal(2009)]{KulkarniKasliwal2009} Kulkarni, S.~R., \& Kasliwal, M.~M.\ 2009, astro2010: The Astronomy and Astrophysics Decadal Survey, 2010, 165

\bibitem[Mason et al.(2010)]{Mason2010} Mason, E., Diaz, M., Williams, R.~E., Preston, G., \& Bensby, T.\ 2010, \aap, 516, A108

\bibitem[Matsuura et al.(2005)]{Matsuura2005} Matsuura, M., Zijlstra, A.~A., Molster, F.~J., Waters, L.~B.~F.~M., Nomura, H., Sahai, R., \& Hoare, M.~G.\ 2005, \mnras, 359, 383

\bibitem[Meaburn et al.(2008)]{Meaburn2008} Meaburn, J., Lloyd, M., Vaytet, N.~M.~H., \& L{\'o}pez, J.~A.\ 2008, \mnras, 385, 269

\bibitem[Mehner et al.(2010)]{Mehner2010} Mehner, A., Davidson, K., Ferland, G.~J., \& Humphreys, R.~M.\ 2010, \apj, 710, 729

\bibitem[Mould et al.(1990)]{Mould1990} Mould, J., et al. 1990, \apjl, 353, L35

\bibitem[Monard(2008)]{Monard2008} Monard, L.~A.~G.\ 2008, \iaucirc, 8946, 1

\bibitem[Nakano et al.(2008)]{Nakano2008} Nakano, S., et al. 2008, \iaucirc, 8972, 1

\bibitem[O'Connor et al.(2000)]{O'Connor2000} O'Connor, J.~A., Redman, M.~P., Holloway, A.~J., Bryce, M., L{\'o}pez, J.~A., \& Meaburn, J.\ 2000, \apj, 531, 336

\bibitem[Ofek et al.(2008)]{Ofek2008} Ofek, E.~O., et al. 2008, \apj, 674, 447

\bibitem[Pastorello et al.(2010)]{Pastorello2010} Pastorello, A., et al. 2010, \mnras, 408, 181

\bibitem[Patat et al.(2010)]{Patat2010} Patat, F., Maund, J.~R., Benetti, S., Botticella, M.~T., Cappellaro, E., Harutyunyan, A., \& Turatto, M.\ 2010, \aap, 510, A108

\bibitem[Peretto et al.(2007)]{Peretto2007} Peretto, N., Fuller, G., Zijlstra, A., \& Patel, N.\ 2007, \aap, 473, 207

\bibitem[Prieto et al.(2008)]{Prieto2008} Prieto, J.~L., et al. 2008, \apjl, 681, L9

\bibitem[Prieto et al.(2009)]{Prieto2009} Prieto, J.~L., Sellgren, K., Thompson, T.~A., \& Kochanek, C.~S.\ 2009, \apj, 705, 1425

\bibitem[Rau et al.(2007)]{Rau2007} Rau, A., Kulkarni, S.~R., Ofek, E.~O., \& Yan, L.\ 2007, \apj, 659, 1536

\bibitem[Reipurth et al.(2002)]{Reipurth2002} {{{  Reipurth, B., Hartmann, L., Kenyon, S.~J., Smette, A., \& Bouchet, P.\ 2002, \aj, 124, 2194 }}}

\bibitem[Sahai et al.(2011a)]{Sahai2011a} Sahai, R., Claussen, M.~J., Schnee, S., Morris, M.~R., \& S{\'a}nchez Contreras, C.\ 2011, (arXiv:1106.4276)

\bibitem[Sahai et al.(2011b)] {Sahai2011b} Sahai, R. , Morris, M.~R., S{\'a}nchez Contreras, C., \&  Claussen, M.~J.\ 2011, ``Understanding the Immediate Progenitors of Planetary Nebulae'', in IAU Symposium 283, ``Planetary Nebulae: an Eye to the Future''

\bibitem[Sahai et al.(1999)]{Sahai1999} Sahai, R., et al.\ 1999, \aj, 118, 468

\bibitem[Sahai et al.(2006)]{Sahai2006} Sahai, R., Young, K., Patel, N.~A., S{\'a}nchez Contreras, C., \& Morris, M.\ 2006, \apj, 653, 1241

\bibitem[Sahai et al.(2003)]{Sahai2003} Sahai, R., Zijlstra, A., S{\'a}nchez Contreras, C., \& Morris, M.\ 2003, \apjl, 586, L81

\bibitem[S{\'a}nchez Contreras et al.(2002)]{Contreras2002} S{\'a}nchez Contreras, C., Desmurs, J.~F., Bujarrabal, V., Alcolea, J., \& Colomer, F.\ 2002, \aap, 385, L1

\bibitem[S{\'a}nchez Contreras et al.(2004)]{Contreras2004} S{\'a}nchez Contreras, C., Gil de Paz, A., \& Sahai, R.\ 2004, \apj, 616, 519

\bibitem[Smith et al.(2009)]{Smith2009} Smith, N., et al. 2009, \apjl, 697, L49

\bibitem[Soker(2000a)]{Soker2000a} {{{  Soker, N.\ 2000a, \aap, 357, 557 }}}

\bibitem[Soker(2000b)]{Soker2000b} Soker, N.\ 2000b, \apj, 540, 436

\bibitem[Soker(2001)]{Soker2001} Soker, N.\ 2001, \mnras, 325, 584

\bibitem[Soker(2007)]{Soker2007} Soker, N.\ 2007, \apj, 661, 490

\bibitem[Soker \& Kashi(2011)]{SokerKashi2011} Soker, N., \& Kashi, A.\ 2011 (arXiv:1107.3454)

\bibitem[Soker \& Tylenda(2006)]{SokerTylenda2006} Soker, N., \& Tylenda, R.\ 2006, \mnras, 373, 733

\bibitem[Szyszka et al.(2009)]{Szyszka2009} Szyszka, C., Walsh, J.~R., Zijlstra, A.~A., \& Tsamis, Y.~G.\ 2009, \apjl, 707, L32

\bibitem[Szyszka et al.(2011)]{Szyszka2011} Szyszka, C., Zijlstra, A.~A., \& Walsh, J.~R.\ 2011, (arXiv:1105.3381)

\bibitem[Thompson et al.(2009)]{Thompson2009} Thompson, T.~A., Prieto, J.~L., Stanek, K.~Z., Kistler, M.~D., Beacom, J.~F., \& Kochanek, C.~S.\ 2009, \apj, 705, 1364

\bibitem[Trammell \& Goodrich(1996)]{Trammell1996} Trammell, S.~R., \& Goodrich, R.~W.\ 1996, \apjl, 468, L107

\bibitem[Tylenda \& Soker(2006)]{TylendaSoker2006} Tylenda, R., \& Soker, N.\ 2006, \aap, 451, 223

\bibitem[Ueta et al.(2007)]{Ueta2007} Ueta, T., Murakawa, K., \& Meixner, M.\ 2007, \aj, 133, 1345

\bibitem[Wagner et al.(2004)]{Wagner2004} Wagner, R.~M., et al.\ 2004, \pasp, 116, 326

\bibitem[Wright et al.(2011)]{Wright2011} Wright, N.~J., Barlow, M.~J., Ercolano, B., \& Rauch, T.\ \mnras, 2011, (arXiv:1107.4554)

\bibitem[Yaron et al.(2005)]{Yaron2005} Yaron, O., Prialnik, D., Shara, M.~M., \& Kovetz, A.\ 2005, \apj, 623, 398


\end{thebibliography}
\end{document}